\documentclass[preprint,showpacs,preprintnumbers,amsmath,amssymb]{revtex4}
\usepackage{graphicx,epsfig,dcolumn,bm,epic,eepic,float}
\usepackage{amsmath}
\usepackage{latexsym}
\usepackage{booktabs}
\usepackage{makeidx,shortvrb,latexsym}
\usepackage{color}
\begin{document}
\title{The EMC ratios of $^4He$, $^3He$ and $^3H$ nuclei in the $k_t$
factorization framework  using the Kimber-Martin-Ryskin unintegrated
parton distribution functions}
\author{M. Modarres }\altaffiliation {Corresponding author, Email:
mmodares@ut.ac.ir, Tel: +98-21-61118645, Fax: +98-21-88004781}
\author{A. Hadian }\altaffiliation {ahmad.hadian@ut.ac.ir}
\affiliation{Physics Department, University of  Tehran, 1439955961,
Tehran, Iran.}
\begin{abstract}
The unintegrated parton distribution functions (UPDFs) of $^3H$,
$^3He$ and $^4He$ nuclei are generated to calculate their structure
functions (SFs)  in the $k_t$-factorization approach. The
Kimber-Martin-Ryskin (KMR) formalisn is applied to evaluate the
double-scale UPDFs of these nuclei from their single-scale parton
distribution functions (PDFs), which can be obtained from the
constituent quark exchange model (CQEM). Afterwards, these SFs are
used to calculate the European Muon Collaboration (EMC) ratios of
these nuclei. The resulting EMC ratios are then compared with the
available experimental data and good agreement with data is
achieved. In comparison with our previous EMC ratios, in which the
conventional PDFs were used in the calculations,   the accord of the
present outcomes with experiment at the small $x$ region becomes
impressive. Therefore, it can be concluded that the $k_t$ dependence
of partons can reproduce the general form of the shadowing effect at
the small $x$ values in above nuclei.
\end{abstract}
\maketitle
\section{Introduction}
The traditional parton distribution functions (PDFs), $a(x,\mu^2)$
($a$=$xq$ and $xg$), depend on the Bjorken variable $x$ (the
longitudinal momentum fraction of the parent hadron) and the squared
scattering factorization scale $\mu^2$. Conventionally, they are
called the integrated PDFs, since the integration over transverse
momentum $k_t$ up to the scale $k_t$ = $\mu$ is performed on them.
Therefore, they are not explicitly depend on the scale $k_t$.
Additionally, these functions are obtained from the global analysis
of deep inelastic and related hard scattering data, and satisfy the
standard Dokshitzer-Gribov-Lipatov-Altarelli-Parisi (DGLAP)
evolution equations \cite{Gribov,Lipatov,Altarelli1,Dokshitzer}.

However, recently, it is observed that unintegrated parton
distribution functions (UPDFs), $f_a(x,k_t^2,\mu^2)$, are necessary
to consider for less inclusive processes, which are sensitive to the
values of transverse momentum of partons. These distributions depend
not only on the factorization scale $\mu$, but also on the
transverse momentum $k_t$. Therefore, they are dependent on two hard
scales, $k_t$ and $\mu$. Application  of the UPDFs to the nuclei,
which was investigated by Martin group \cite{Oliveira1}, have
demonstrated that it can significantly affect the nucleus structure
function (SF) at the small $x$ region. In addition, very recently,
we illustrate that especially at the small Bjorken values ($x\ll
0.1$), the UPDFs have an enormous effect on the SF and European Muon
collaboration \cite{Aubert1} (EMC) ratio of $^6Li$ nucleus
\cite{Hadian4} which is known as shadowing effect
\cite{Frankfurt1,Frankfurt2}. Due to dependency of the UPDFs on the
extra hard scale $k_t$, compared with the usual PDFs, we potentially
have to deal with the much more complicated
Ciafaloni-Catani-Fiorani-Marchesini (CCFM) evolution equations
\cite{Ciafaloni,Catani,Catani2,Marchesini,Marchesini2}.

Working with the CCFM equations, of course, confront two major
problems. First, practically, these equations are used only in the
Monte Carlo event generators
\cite{Marchesini3,Marchesini4,Jung,Jung2,Jung3}, and so solving them
is a mathematically complicated task. Second, these kind of
equations are incapable to generate a complete quark version and can
be exclusively used for the gluon contributions
\cite{Ciafaloni,Catani,Catani2,Marchesini,Marchesini2}. Therefore,
to overcome these obstacles, Kimber, Martin and Ryskin (KMR)
introduced the more efficient $k_t$-factorization framework
\cite{Kimber1,Kimber2,kimber3}. The KMR approach was constructed
around the standard LO DGLAP evolution equations, along with a
modification due to the angular ordering condition (AOC), which is
the essential dynamical property of the CCFM formalism. This
prescription was successfully applied by us to investigate different
hard scattering processes in the various studies; e.g. see the
references
\cite{Modarres2,Modarres3,Modarres4,Modarres5,Modarres6,Modarres7,Modarres8,Modarres9,Modarres10,
Modarres11,Olanj1,Hosseinkhani1}. In the section 3, we briefly
introduce this approach as a method to generate the double-scale
UPDFs from the conventional single-scale PDFs.

To generate the UPDFs by using the KMR procedure, the integrated
PDFs are required as inputs. So, we use the constituent quark
exchange model (CQEM) to obtain the PDFs of $^4{He}$, $^3He$ and
$^3H$  nuclei at the hadronic scale $\mu_0^2$ = 0.34 $GeV^2$
\cite{Hadian3,Rasti,Zolfagharpour}. These resulting PDFs at the
initial scale $\mu_0^2$, are then evolved to any required higher
energy scale $Q^2$ by using the standard DGLAP evolution equations
\cite{Botje}. We will discuss about this process in the section 2.

So, in what follows, first in the section 2, based on the CQEM, the
PDFs of  $^4{He}$, $^3He$ and $^3H$  will be calculated. The
sections 3 contains a brief introduction to the KMR formalism and
the formulation of SF ($F_2(x,Q^2)$) in the $k_t$ factorization
framework. Finally,  results, discussions  and conclusion are
presented in the section 4.
\section{The PDF$s$ of  the $^4{He}$, $^3He$ and $^3H$  nuclei in the CQEM}
In this section, we tend to obtain the point-like valence quark, sea
quark and gluon distributions of $^4{He}$, $^3He$ and $^3H$  nuclei.
To reach our purpose, the CQEM, which indeed consists of two more
basic schemes, is applied. These two primary approaches are the
quark exchange framework (QEF) \cite{Jaffe1,Hoodbhoy1} and the
constituent quark model (CQM) \cite{Feynman,Close,Roberts}. The QEF
was first suggested by Hoodbhoy and Jaffe to calculate the valence
quark momentum distributions of $A=3$ iso-scalar system
\cite{Jaffe1,Hoodbhoy1}, and afterwards, was successfully
reformulated by us for the $^6Li$ and $^4He$ nuclei
\cite{Hadian3,Hadian1,Hadian2}. However, this approach is unable to
generate the other partonic degrees of freedom, i.e., the sea quarks
and the gluons. To consider these extra distributions, the CQM,
which was first introduced by Feynman \cite{Feynman,Close,Roberts},
is incorporated in the QEF. This combination, like our previous
works (e.g. references \cite{Hadian2,Hadian3,Rasti,Hadian4}), is
denominate the CQEM (=QEF $\oplus$ CQM).

The up and down constituent quark momentum distributions of $^3He$
and $^3H$ nuclei, which were calculated by using the QEF in the
reference \cite{Zolfagharpour}, can be written as follows:
\begin{equation}\label{1}
\rho_{\mathcal{U}}^{^3He}(k)=\rho_{\mathcal{D}}^{^3H}(k)=\Big[{{{2}A(k)+{2\over9}B(k)-{16\over27}C(k)+{28\over27}D(k)}\Big]
{\Big[1+{9\over8}\mathcal{I}\Big]}}^{-1},
\end{equation}
\begin{equation}\label{2}
\rho_{\mathcal{D}}^{^3He}(k)=\rho_{\mathcal{U}}^{^3H}(k)=\Big[{{A(k)+{1\over9}B(k)-{20\over27}C(k)+{26\over27}D(k)}\Big]
{\Big[1+{9\over8}\mathcal{I}\Big]}}^{-1},
\end{equation}
where $\rho_{\mathcal{U}}$ and $\rho_{\mathcal{D}}$ represent the up
and down constituent quark momentum distributions, respectively. For
the $^4He$ iso-scalar nucleus, the up and down  momentum
distributions are equal, and these distributions, which were
computed in the reference \cite{Hadian3}, can be presented as
follows:
\begin{equation}\label{3}
\rho_{\mathcal{U}}^{^4He}(k)=\rho_{\mathcal{D}}^{^4He}(k)=\Big[{{{6}A(k)+{2}B(k)+{4\over3}C(k)+{2\over3}D(k)}\Big]
{\Big[1+{9\over4}\mathcal{I}\Big]}}^{-1}.
\end{equation}
In the above equations, the coefficients $A, B, C, D,$ and the
overlap integral $\mathcal{I}$  are defined as follows:
\begin{equation}\label{4}
A(k)=\Big({3b^2\over2\pi}\Big)^{3\over2}exp{\Big[-{3\over2}b^2{k^2}\Big]},
\end{equation}
\begin{equation}\label{5}
{\quad}B(k)=\Big({27b^2\over8\pi}\Big)^{3\over2}exp{\Big[-{3\over2}b^2{k^2}\Big]}\mathcal{I},
\end{equation}
\begin{equation}\label{6}
{\quad}{\quad}C(k)=\Big({27b^2\over7\pi}\Big)^{3\over2}exp{\Big[-{12\over7}b^2{k^2}\Big]}\mathcal{I},
\end{equation}
\begin{equation}\label{7}
{\quad}D(k)=\Big({27b^2\over4\pi}\Big)^{3\over2}exp{\Big[-{3}b^2{k^2}\Big]}\mathcal{I}.
\end{equation}
\begin{align}\label{8}
\mathcal{I}=8\pi^2\int^{\infty}_{0}{x^2dx}\int^{\infty}_{0}{y^2dy}\int^{1}_{-1}d(cos{\theta})
{exp{\Big[-{{{3}x^2}\over{4b^2}}\Big]}}{\mid{\chi(x,y,cos{\theta})}\mid}^2,
\end{align}
where $\chi$ is the nuclear wave function and parameter $b$ is the
nucleon's radius. Note that the basic expressions in this section
are based on the naive harmonic oscillator model for the constituent
quarks. In the present study,  we intend to concentrate only on the
pure quark-exchange effect, dynamically. Therefore, to reduce the
number of variables, we suppose the  same nucleon’s radius,  $b$ =
0.8 $fm$,  for the $^4He$, $^3He$  and $^3H$ nuclei, with
corresponding overlap integral $\mathcal{I}$. The thorough
discussions about calculating the above momentum distributions for
the $^4He$ and $^3He$ nuclei in the QEF, were given in the
references \cite{Hadian3} and \cite{Zolfagharpour}, respectively.
Now, the constituent quark distributions in the nucleons of the
nucleus $\mathcal{A}_i$, at each $Q^2$, can be related to the above
momentum distributions, as follows ($j$ = $p$, $n$ ($a$ =
${\mathcal{U}}$, ${\mathcal{D}}$) for the proton (up quark) and
neutron (down quark), respectively) \cite{Jaffe1}:
\begin{equation}\label{9}
f^j_a(x,Q^2;\mathcal{A}_i)=\int{{{\rho^j_a(\vec{k};\mathcal{A}_i)}}\delta\Big({x-{k_+\over
M}}\Big)d{\vec{k}}},
\end{equation}
the reason for the $Q^2$ dependence of the right hand side of the
equation (9) will be explained below. The light-cone momentum of the
constituent quark in the target rest frame is used and $k^0$ is
considered as a function of ${|\vec{k}|}$
${({k^0=[{({{\vec{k}^2}+m_a^2}})}^{1\over2}-{\epsilon}_0^a})$. The
two free parameters, i.e., ${m_a}$ and ${\epsilon}_0^a$, are the
quark masses and their binding energies, respectively. We can
determine these free parameters such that the best fit to the
valence quark distribution functions of Martin $et$ $al.$, i.e.,
MSTW 2008 \cite{Stirling,Stirling2,Stirling3}, is achieved, at
$Q^2=0.34$ $GeV^2$. By doing so, for the $^4He$, the pair of ($m_a$,
$\epsilon_0^a$) is chosen as (320, 120 MeV) ($a$ = ${\mathcal{U}}$,
${\mathcal{D}}$), and for the $^3He$,  the pairs of
($m_{\mathcal{U}}$, $\epsilon_0^{\mathcal{U}}$) and
($m_{\mathcal{D}}$, $\epsilon_0^{\mathcal{D}}$) are taken as (300,
130 MeV) and (325, 115 MeV) (they will be interchanged for  $^3H$),
respectively. After doing the angular integration, the equation
(\ref{9}) leads to the following constituent quark distributions:
\begin{equation}\label{10}
f^a_j(x,Q^2;\mathcal{A}_i)={{2\pi
M}}\int^{\infty}_{k^a_{min}}{{{\rho^a_j(\vec{k};\mathcal{A}_i)}}kdk},
\end{equation}
with,
\begin{equation}\label{11}
k^a_{min}(x)={({{x M}+{\epsilon^a_0})^2}-m^2_a\over{2({x
M}+{\epsilon^a_0})}},
\end{equation}
where $M$ indicates the nucleon mass. Because of the above fitting
the right hand side of the equations (9) and (10) become $Q^2$
dependent.

By determination of the constituent distributions of $^4He$, $^3He$
and $^3H$ nuclei via the QEF, it's the time to present a brief
description of the CQM to complete our discussion about the CQEM. In
the CQM, it is supposed that the constituent quarks are not
fundamental objects, but instead consist of point-like partons
\cite{Feynman,Close,Roberts}. Therefore, their structure functions
can be expressed by a set of functions, $\phi_{ab}(x)$, which define
the number of partons of type $b$ inside the constituent of type $a$
with the fraction $x$ of its total momentum. The various types and
functional forms of the constituent quarks structure functions are
extracted from  three natural assumptions, namely: (i) the
determination of the point-like partons by QCD, (ii) the Regge
behavior for x $\rightarrow$ 0 as well as the duality idea, and,
(iii) the isospin and the charge conjugate invariant. For different
kinds of partons, the following definitions of the structure
functions have been proposed: in the case of valence quarks,
\begin{equation}\label{5p}
\phi_{\mathcal{P}q_v}\Big({x\over z},
\mu_0^2\Big)={{\Gamma(A+{1\over2})}\over{\Gamma({1\over2})\Gamma(A)}}{{\Big({1-{x\over
z}\Big)}^{A-1}} \over \sqrt{x\over z}},
\end{equation}
for the sea quarks,
\begin{equation}\label{6p}
\phi_{\mathcal{P}q_s}\Big({x\over z}, \mu_0^2\Big)={C\over{x\over
z}}{{\Big({1-{x\over z}\Big)}^{D-1}}},
\end{equation}
and finally,  for the gluons,
\begin{equation}\label{7p}
\phi_{\mathcal{P}g}\Big({x\over z}, \mu_0^2\Big)={G\over{x\over
z}}{{\Big({1-{x\over z}\Big)}^{B-1}}}.
\end{equation}
The momentum carried by the second moments of the parton
distributions are known experimentally at high $Q^2$. Their values
at the low scale $Q_0^2$ could be obtained by performing a
next-to-leading-order evolution downward. These procedure is used to
extract the value of the constants $A, B, G$ and the ratio $C$/$D$.
For example, at the hadronic scale ${Q_0}^2$ = 0.34 $GeV^2$, 53.5
  of the nucleon momentum is carried by the valence quarks, 35.7
by the gluons and the remaining momentum are belong to the sea
quarks. So, in this scale, the mentioned parameters take the
following values: $A=0.435$, $B=0.378$, $C=0.05$,  $D=2.778$ and
$G=0.135$. More information and detailed discussion about the above
structure functions for different kinds of partons, and the
procedures of evaluating these constants can be found in the
references
\cite{Altarelli,Scopetta1,Manohar,Vento,Rasti,Yazdanpanah2}.
Ultimately, the main equation of the CQM can be written as follows:
\begin{equation}\label{12}
q(x,\mu_0^2)=\int^{1}_{x}{dz\over
z}\Big[{\mathcal{U}(z,\mu_0^2)\phi_{\mathcal{U}q}\Big({x\over z},
\mu_0^2\Big)+\mathcal{D}(z,\mu_0^2)\phi_{\mathcal{D}q}\Big({x\over
z}, \mu_0^2\Big)}\Big],
\end{equation}
where $q$ denotes the various point-like partons, i.e., valence
quarks ($u_v$, $d_v$), sea quarks ($u_s$, $d_s$, $s$), sea
anti-quarks ($\bar{u}_s$, $\bar{d}_s$, $\bar{s}$) and gluons ($g$).
The $\mathcal{U}$ and $\mathcal{D}$ indicate the distributions of up
and down constituent quarks, respectively. Actually, these
quantities are the same as the functions $f_j^a$ ($a=\mathcal{U}$,
$\mathcal{D}$) of the equation (10), and for simplicity, since then,
we replace the $f_j^{\mathcal{U}}$ and $f_j^{\mathcal{D}}$ labels by
$\mathcal{U}$ and $\mathcal{D}$, respectively. The $\mu_0^2$ = 0.34
$GeV^2$ is the initial hadronic scale at which the CQM is defined.
In the CQM, the sea quark and anti-quark distributions are
independent of iso-spin flavor. Therefore, in the following, the
label $q_s$ represents both sea quark and anti-quark distributions.
It should be noted that, the structure functions
$\phi_{\mathcal{U}d}\Big({x\over z},\mu_0^2\Big)$ and
$\phi_{\mathcal{D}u}\Big({x\over z}, \mu_0^2\Big)$ in the equation
($\ref{12}$) are zero, because  in the constituent quark of type
$\mathcal{U}$, there is no point-like valence quark of type $d$ and
vice versa (see the reference \cite{Altarelli} about the origin of
this assumption). In addition, for the $^4{He}$ nucleus, the
constituent up and down quark distributions are equal, because
unlike the $^3He$ and $^3H$ cases, it is an iso-scalar system.

Therefore, eventually, the single-scale PDFs of $^4He$, $^3He$ and
$^3H$ nuclei at the hadronic scale $\mu_0^2$ can be specified in the
CQEM as follows:
\\(i) for the  $^4He$ nucleus,
\begin{equation}\label{13}
u_v^{^4He}(x,\mu_0^2)=d_v^{^4He}(x,\mu_0^2)=\int^{1}_{x}{dz\over
z}{\mathcal{U}^{^4He}(z,\mu_0^2)\phi_{\mathcal{U}q_v}\Big({x\over
z}, \mu_0^2\Big)},
\end{equation}
\begin{equation}\label{14}
q_s^{^4He}(x,\mu_0^2)=2\int^{1}_{x}{dz\over
z}{\mathcal{U}^{^4He}(z,\mu_0^2)\phi_{\mathcal{U}q_s}\Big({x\over
z}, \mu_0^2\Big)},
\end{equation}
\begin{equation}\label{15}
g^{^4He}(x,\mu_0^2)=2\int^{1}_{x}{dz\over
z}{\mathcal{U}^{^4He}(z,\mu_0^2)\phi_{\mathcal{U}g}\Big({x\over z},
\mu_0^2\Big)},
\end{equation}
where
\begin{equation}\label{16}
\mathcal{U}^{^4He}(z,\mu_0^2)={{2\pi
M}}\int^{\infty}_{k_{min}}{{{\rho_{\mathcal{U}}^{^4He}(k)}}kdk},
\end{equation}
\\(ii) for the $^3He$ and $^3H$ nuclei,
\begin{equation}\label{17}
u_v^{^3He}(x,\mu_0^2)=d_v^{^3H}(x,\mu_0^2)=\int^{1}_{x}{dz\over
z}{\mathcal{U}^{^3He}(z,\mu_0^2)\phi_{\mathcal{U}u_v}\Big({x\over
z}, \mu_0^2\Big)},
\end{equation}
\begin{equation}\label{18}
d_v^{^3He}(x,\mu_0^2)= u_v^{^3H}(x,\mu_0^2) = \int^{1}_{x}{dz\over
z}{\mathcal{D}^{^3He}(z,\mu_0^2)\phi_{\mathcal{D}d_v}\Big({x\over
z}, \mu_0^2\Big)},
\end{equation}
\begin{equation}\label{19}
q_s^{^3He}(x,\mu_0^2)= q_s^{^3H}(x,\mu_0^2) = \int^{1}_{x}{dz\over
z}\Big[{\mathcal{U}^{^3He}(z,\mu_0^2)\phi_{\mathcal{U}q_s}\Big({x\over
z},
\mu_0^2\Big)}+{\mathcal{D}^{^3He}(z,\mu_0^2)\phi_{\mathcal{D}q_s}\Big({x\over
z}, \mu_0^2\Big)}\Big],
\end{equation}
\begin{equation}\label{20}
g^{^3He}(x,\mu_0^2)= g^{^3H}(x,\mu_0^2) = \int^{1}_{x}{dz\over
z}\Big[{\mathcal{U}^{^3He}(z,\mu_0^2)\phi_{\mathcal{U}g}\Big({x\over
z},
\mu_0^2\Big)}+{\mathcal{D}^{^3He}(z,\mu_0^2)\phi_{\mathcal{D}g}\Big({x\over
z}, \mu_0^2\Big)}\Big],
\end{equation}
where
\begin{equation}\label{21}
\mathcal{U}^{^3He}(z,\mu_0^2)={{2\pi
M}}\int^{\infty}_{k_{min}}{{{\rho_{\mathcal{U}}^{^3He}(k)}}kdk},
\end{equation}
and
\begin{equation}\label{22}
\mathcal{D}^{^3He}(z,\mu_0^2)={{2\pi
M}}\int^{\infty}_{k_{min}}{{{\rho_{\mathcal{D}}^{^3He}(k)}}kdk}.
\end{equation}
These resulted PDFs for the $^4He$ and $^3He$ nuclei, at the
hadronic scale $\mu_0^2$ = 0.34 $GeV^2$, are shown in the panels (a)
and (b) of figure 1, respectively.

Now, by using the  standard DGLAP equations, the above PDFs which
are obtained from the CQEM at the initial scale  $\mu_0^2$, can be
evolved to any higher energy scale $Q^2$ \cite{Botje}. However,
these conventional PDFs are not $k_t$-dependent distributions. So,
to consider the transverse momentum explicitly, in the next section
the KMR approach will be introduced to generate the double-scale
UPDFs from these single-scale PDFs.

\section{The KMR formalism and the UPDF$s$ and SF calculations}
It is well known that there are problems at small $x$ region
\cite{m1,m2,m3,m4}. So one should use the general formalism such as
CCFM which the transverse momentum of partons play the crucial role
or the reggeon theory such as pameron model. However it was shown
that the $k_t$-factorization formalism is capable to consider the
precise kinematics of the process and an important part of the
virtual loop corrections, via the survival probability factor $T$
(see below). On the other hand, if we work with integrated partons,
we have to include the NLO (and sometimes the NNLO) contributions to
account for these effects. These differences appear to cause a
discrepancy between the integrated and unintegrated frameworks
\cite{Kimber2,Kimber1,kimber3}.

A brief description of the KMR formalism as well as the SF formula
in the $k_t$-factorization framework, is presented in the following
subsections (A and B), respectively.
\subsection{The KMR formalism}
 In this subsection, we   briefly discuss about the KMR scheme to
 extract the UPDFs from the resulted integrated PDFs of the previous section, as inputs. The KMR
 formalism was first proposed by Kimber, Martin and Ryskin \cite{Kimber2,Kimber1,kimber3}.
 From the two scheme discussed in the
reference \cite{Kimber2} we use the second approach which directly
relates the UPDFs to the conventional  PDFs. This formalism was also
separately discussed in the reference \cite{kimber3}.
 Based on this scheme, the  $LO$ DGLAP equations can be modified  by separating
 the real and virtual contributions of the evolution, and  the two-scale UPDFs,
 $f_a(x,k_t^2,\mu^2)$ ($a$ = $q$ or $g$), can be defined as follows:
\begin{equation}\label{23}
f_a(x,k_t^2,\mu^2) = T_a(k_t^2,\mu^2)\sum\limits_{b=q,g}\Big[
{\alpha_s(k_t^2)\over{2\pi}}\int^{1-\Delta}_{x}dzP_{ab}^{(LO)}(z)b\Big({x\over
z},k_t^2\Big)\Big],
\end{equation}
where $P_{ab}^{(LO)}$ represent the $LO$ splitting functions, which
account for the probability of a parton of type $a$ with momentum
fraction $x^{\prime\prime}$, $a(x^{\prime\prime},Q^2)$, emerging
from a parent parton of type $b$ with a larger momentum fraction
$x^\prime$, $b(x^\prime,Q^2)$, through $z =
x^{\prime\prime}/x^\prime$. The survival probability factor, i.e.,
Sudakov form factor $T_a$, which gives the probability that parton
$a$ with transverse momentum $k_t$ remains untouched in the
evolution up to the factorization scale $\mu$, is defined via the
following equation:
\begin{equation}\label{24}
T_a(k_t^2,\mu^2)=exp\Big(-\int^{\mu^2}_{k_t^2}{\alpha_s(k^2)\over{2\pi}}{dk^2\over
k^2}\sum\limits_{b=q,g}\int^{1-\Delta}_{0}{dz^\prime}P_{ab}^{(LO)}(z^\prime)\Big),
\end{equation}
The infrared cut-off, $\Delta$ = $1-z_{max}$ = $k_t/(\mu+k_t$), is
determined by imposing the AOC on the last step of the evolution,
and protects the $1/(1-z)$ singularity in the splitting functions
arising from the soft gluon emission. In the KMR formulation, the
key idea is that the dependence on the second scale $\mu$ of the
UPDFs appears only at the last step of the evolution. By completing
the procedures of producing the UPDFs from the KMR scheme, the UPDFs
of the $^4He$, $^3He$ and $^3H$ nuclei  can be evaluated by using
their conventional PDFs (which were determined in the previous
section), as inputs.
\subsection{The SF in the $k_t$-factorization framework}
Here it is briefly described the different steps to calculate  the
SF ($F_2(x,Q^2)$) in the $k_t$-factorization framework,  by using
the KMR UPDFs as inputs. We explicitly investigate the separate
contributions of gluons and (direct) quarks to the SF expression
\cite{Kimber2,Kimber1,kimber3}.

The unintegrated gluons can contribute to $F_2$ via an intermediate
quark.  As shown in the figure 2, both the quark box and crossed-box
diagrams must be regarded as the gluon portions. The variable $z$
denotes the fraction of the gluon's momentum that is transferred to
the exchanged struck quark. The parameters $k_t$ and $\kappa_t$
indicate the transverse momentum of the parent gluons and daughter
quarks, respectively. In the $k_t$-factorization framework, the
unintegrated gluon contributions to $F_2$ can be obtained via the
following equation
\cite{Kimber1,Kimber2,kimber3,Kwiecinski,Askew,Stasto}:
\begin{align}\label{25}
F_2^{g\rightarrow q\bar{q}}(x&,Q^2)=\sum_q\:e_q^2{Q^2\over
4\pi}\int{dk_t^2\over {k_t^4}}\int_0^1 d\beta\int
d^2\kappa_t\:\alpha_s(\mu^2)\:f_g\Big({x\over
z},k_t^2,\mu^2\Big)\Theta\Big(1-{x\over
z}\Big)\nonumber\\&\times\Big\lbrace[\beta^2+{(1-\beta)}^2]\:\Big({\boldsymbol{\kappa}_t
\over D_1}-{{\boldsymbol{\kappa}_t-\boldsymbol{k}_t}\over
D_2}\Big)^2\:+\:[m_q^2+4Q^2\beta^2(1-\beta)^2]\Big({1\over
D_1}-{1\over D_2}\Big)^2\Big\rbrace.
\end{align}
The variable $\beta$ is defined as the light-cone fraction of the
photon's momentum carried by the internal quark line. In addition,
the denominator factors are defined as follows:
\begin{align}\label{26}
&D_1=\kappa_t^2+\beta(1-\beta)Q^2+m_q^2\nonumber\\&D_2=(\boldsymbol{\kappa}_t-\boldsymbol{k}_t)^2+\beta(1-\beta)Q^2+m_q^2.
\end{align}
In the equation (\ref{25}), the summation goes over various quark
flavors $q$ with different masses $m_q$ which can appear in the box.
In the present study, we consider three lightest flavor of quarks
($n_f$ = 3), i.e., $u$, $d$ and $s$, whose masses are neglected with
a good approximation. So, $n_f$ = 3 throughout of our calculations.
Additionally, the variable $z$  is defined as follows:
\begin{equation}\label{27}
{1\over
z}=1\:+\:{{\kappa_t^2+m_q^2}\over{(1-\beta)Q^2}}\:+\:{{k_t^2+\kappa_t^2-2\:\boldsymbol{\kappa}_t\:.\:\boldsymbol{k}_t+m_q^2}\over
{\beta Q^2}}.
\end{equation}
which is the ratio of the Bjorken variable $x$ and the fraction of
the proton momentum carried by the gluon. As in the reference
\cite{Kwiecinski}, the scale $\mu$, which controls the unintegrated
gluon distribution and the QCD coupling constant $\alpha_s$, is
chosen as follows:
\begin{equation}\label{28}
\mu^2=k_t^2+\kappa_t^2+m_q^2.
\end{equation}
The equation (\ref{25}) gives the contributions of   unintegrated
gluons to $F_2$ in the perturbative region, $k_t$ $>$ $k_0$, where
the UPDFs are defined. The smallest cutoff, $k_0$, we can choose, is
the initial scale of order 1 $GeV$, at which the $k_t$-factorization
scheme is defined \cite{Askew}. For the contribution from the
nonpertubative region, $k_t$ $<$ $k_0$, it can be approximated
\begin{align}\label{29}
\int_0^{k_0^2}{dk_t^2 \over
k_t^2}\:f_g(x,k_t^2,\mu^2)\:\Big[{\textit{remainder\:\:of\:\:equation}\:\:(\ref{25})\over
k_t^2}\Big]\simeq\:xg(x,k_0^2)\:T_g(k_0,\mu)\:\Big[\quad\Big]_{k_t=a},
\end{align}
where $a$ is belong to the interval (0, $k_0$). The dependence on
the choice of $a$  is numerically unimportant to the nonperturbative
contribution \cite{Kimber1,Kimber2,kimber3}.

Now, the contributions of unintegrated quarks must be added to
$F_2$. If an initial quark with Bjorken scale $x$/$z$ and
perturbative transverse momentum $k_t$ $>$ $k_0$, splits to a
radiated gluon and a quark with smaller Bjorken scale $x$ and
transverse momentum $\kappa_t$, this final quark can then couple to
the photon and contributes to $F_2$, as follows:
\begin{align}\label{30}
F_2^{q(perturbative)}(x,Q^2)=\sum_{q=u,d,s} \:e_q^2
\:\int_{k_0^2}^{Q^2}\:&{d\kappa_t^2\over
\kappa_t^2}\:{\alpha_s(\kappa_t^2)\over{2\pi}}\int_{k_0^2}^{\kappa_t^2}\:{dk_t^2\over
k_t^2}\int_x^{Q/{(Q+k_t)}}\:dz\nonumber\\&\times\Big[f_q
\Big({x\over z},k_t^2,Q^2 \Big)\:+\:f_{\bar{q}}\: \Big({x\over
z},k_t^2,Q^2 \Big)\Big]P_{qq}(z),
\end{align}
where   during the quark evolution, AOC is imposed on the upper
limit of the $z$ integration. Again, one must consider  the
nonperturbative contributions for the $k_t$ $<$ $k_0$,
\begin{equation}\label{31}
F_2^{q(nonperturbative)}(x,Q^2)=\sum_q
e_q^2\:\Big(xq(x,k_0^2)\:+\:x\bar{q}(x,k_0^2)\Big)\:T_q(k_0,Q),
\end{equation}
which physically can be assumed as a quark or anti-quark, which does
not experience real splitting in the perturbative region, and
interacts unchanged, with the photon at the scale $Q$. So, a
Sudakov-like factor, $T_q(k_0, Q)$, is written to indicate the
probability of evolution from $k_0$ to $Q$ without radiation.

Finally, by   summing  both gluon and quark contributions, one can
obtain the overall SF in the $k_t$-factorization framework.
Subsequently, the EMC ratio, which is defined as the ratio of the SF
of the bound nucleon to that of the free nucleon, can be evaluated
as follows \cite{Jaffe1}:
\begin{equation}\label{32}
\mathcal{R}_{EMC}={F_2^T(x)\over{F^{T^\star}_2(x)}},
\end{equation}
where $T$ stands for the target, averaged over nuclear spin and
iso-spin and $T^\star$ is a hypothetical target with exactly the
same quantum numbers but  with no  parton exchange \cite{Jaffe1}.
So, if the overlap integral $\mathcal{I}$ is omitted in the momentum
distribution formula, i.e., the equations (\ref{1})-(\ref{3}), we
can compute the SF of free nucleons. The effects of nuclear Fermi
motion are neglected from both $T$ and $T^\star$. We utilize the KMR
UPDFs to calculate the SFs, and the EMC ratios of $^4He$, $^3He$ and
$^3H$ nuclei in the $k_t$ factorization approach, which will be
presented in the next section.
\section{Results, discussions, and conclusions}
The overall SF, $F_2$, of  $^4He$ nucleus in the $k_t$-factorization
framework, using the KMR UPDFs, at the energy scales $Q^2$ = 4.5 and
27 $GeV^2$ are plotted in the panels (a) and (b) of Figure 3,
respectively (the full curves). As expected,  by increasing the
scale $Q^2$ from 4.5 to 27 $GeV^2$, a considerable rise in $F_2$ at
the smaller values of $x$ occurs. The dash curves in each panel, are
the SFs of the free proton in the $k_t$-factorization framework, in
which to generate the KMR UPDFs, the MSTW 2008 PDF sets are used as
inputs. The SF of a hypothetical $^4He$ target, without any quark
exchange between its nucleons (by ignoring the overlap integral
$\mathcal{I}$ in the momentum density equation), i.e., the
hypothetical free nucleon,  in the $k_t$-factorization framework
using the KMR UPDFs, are also exhibited in this figure for
comparison (the dotted curves). The three lightest flavors of
quarks, i.e., $u$, $d$ and $s$, are considered in calculation of
these SFs. According to the equation (\ref{32}), the EMC ratio in
the  $k_t$-factorization formalism at each energy scale, can be
evaluated by regarding the ratio of the full curve ($^4He$ SF) to
the dotted curve (the hypothetical free nucleon SF). It is observed
that the SFs of our hypothetical free nucleon are in overall  good
agreement with the SF of the free proton. Especially at the small
$x$ region, as one should expect, the SFs of free nucleon (the
dotted curves) and free proton (the dash curves) are approximately
equal, since in this area, $u$ = $d$ = $\bar u$ = $\bar d$ and the
proton and neutron SFs must be the same. The similar conclusions
have been made for the $^6Li$ nucleus in our recent work, i.e., the
reference \cite{Hadian4}.

Figures 4 and 5 are the same as figure 3, but for the $^3He$ and
$^3H$ nuclei, respectively. Similar to the figure 3,  the SFs of
free nucleon (the dotted curves) and free proton (the dash curves)
are again approximately equal at the small $x$. Also, to obtain the
EMC ratio in the  $k_t$-factorization framework for these nuclei,
one should  again consider the ratio of the full curves to the
dotted curves in each panel. In addition, as we increase the $Q^2$
value to 27 $GeV^2$, again, the overall SFs become greater.

The resulting EMC ratios of $^4He$, $^3He$ and $^3H$ nuclei in the
$k_t$-factorization framework, using the KMR UPDFs,  are plotted in
the figures 6, 7 and 8, respectively. For each of  these nuclei, the
ratio is calculated at the energy scales  $Q^2$ = 4.5 and 27
$GeV^2$. Due to neglecting the Fermi motion, the EMC ratios
monotonically decrease and the growth in the EMC ratios at the large
$x$ values do not occur. Therefore, the EMC ratios are illustrated
for the $x \le 0.7$ region.   In the figure 6, the experimental
measurements are from the JLab \cite{Seely,Malace} (filled circles),
NMC  \cite{New,Malace} (filled triangles), and SLAC
\cite{Gomez,Malace} (filled squares), while in the figures 7 and 8,
the filled circles and the filled squares are the experimental data
from JLab \cite{Seely,Malace} and HERMES \cite{Malace,Airapetisn},
respectively. To compare the theoretical and experimental $^4He$ EMC
ratios more clearly at the small $x$, the experimental NMC data are
illustrated in the distinct diagrams with logarithmic scale, i.e.,
panels (b) and (d) of the figure 6. The dash curves in the panels
(a) and (b) of the figure 6, are given from our prior work
\cite{Hadian3}, in which the $k_t$ dependence of parton distribution
functions were neglected in the $^4He$ EMC calculations. Obviously,
at the small $x$ region, the present  $^4He$ EMC results are
extremely improved with respect to our previous outcomes
\cite{Hadian3}. However, when the Bjorken scale $x$ is increased,
the differences between the full and dash curves decrease, which
show that the $k_t$-factorization scheme has an important effect on
the  EMC calculations at the small $x$ values
\cite{Hadian4,Frankfurt1,Frankfurt2}, i.e., shadowing region.
Therefore, the inclusion of $k_t$-dependent PDFs in the EMC
calculation, can reproduce the general form of shadowing effect
\cite{Hadian4,Frankfurt1,Frankfurt2}. The similar behavior is seen
in the EMC curves of $^3He$ and $^3H$ nuclei (see the figures 7 and
8, respectively) as well as the  EMC ratio of $^6Li$ nucleus (see
the figure 10 of reference \cite{Hadian4}). In addition, for all
three nuclei which discussed here, the EMC curves at the energy
scales 4.5 and 27 $GeV^2$ have approximately the same behavior (see
also the EMC ratio of $^6Li$ nucleus in the figure 10 of reference
\cite{Hadian4}). This similarity is expected, because the EMC ratio
are not $Q^2$ dependent, significantly (e.g. see the reference
\cite{Gomez}).

The comparisons of EMC ratios of $^6Li$ (the dash-dotted curves),
$^4He$ (the dash curves), $^3He$ (the dotted curves) and $^3H$ (the
full curves) nuclei in the KMR approach at the energy scales 4.5 and
27 $GeV^2$ are displayed in the left and right panels of figure 9,
respectively. The  $^6Li$  EMC ratios are plotted from the reference
\cite{Hadian4}. As expected, the EMC curves of $^3He$ and $^3H$
mirror nuclei are very close together, because of iso-spin symmetry
assumption. However, by increasing the number of nucleons in the
nucleus, the probabilities of quark exchanges among the nucleons are
  increased, which make the $\mathcal{R}_{EMC}$ to have greater
deviation from unity.

In conclusion, the CQEM and the KMR UPDFs were used to obtain the
EMC ratios of $^4He$, $^3He$ and $^3H$ nuclei in the
$k_t$-factorization framework. To calculate the double-scale UPDFs,
we needed the conventional single-scale PDFs for each nucleon as
inputs. Therefore, the CQEM was employed to elicit the PDFs of these
nuclei at the hadronic scale 0.34 $GeV^2$. Then, the resulted PDFs
were
 evolved by the DGLAP evolution equations to the higher energy
scales. Subsequently, by using the KMR UPDFs, the SFs of  these
nuclei in the $k_t$-factorization scheme were calculated at the
energy scales $Q^2$ = 4.5 $GeV^2$ and 27 $GeV^2$. Subsequently, we
compared the resulted  SFs with the corresponding SF of free proton.
Eventually, after computing the EMC ratios of $^4He$, $^3He$ and
$^3H$ nuclei, they were compared with the experimental data. It was
seen that the outcome EMC ratios astonishingly were consistent with
the various experimental data. Especially, the $k_t$-factorization
approach extremely improved the EMC ratios of mentioned nuclei at
the shadowing region. Therefore, similar to our previous work
\cite{Hadian4}, the reduction of EMC effect at the small $x$ region,
which traditionally is  known as the "shadowing phenomena"
\cite{Frankfurt1,Frankfurt2}, can be successfully explained in the
$k_t$-factorization framework by using the KMR UPDFs.
\section*{Acknowledgements}
$MM$ would like to acknowledge the Research Council of University of
Tehran for the grants provided for him.

\begin{figure}[b!]
  \includegraphics [width=\linewidth]{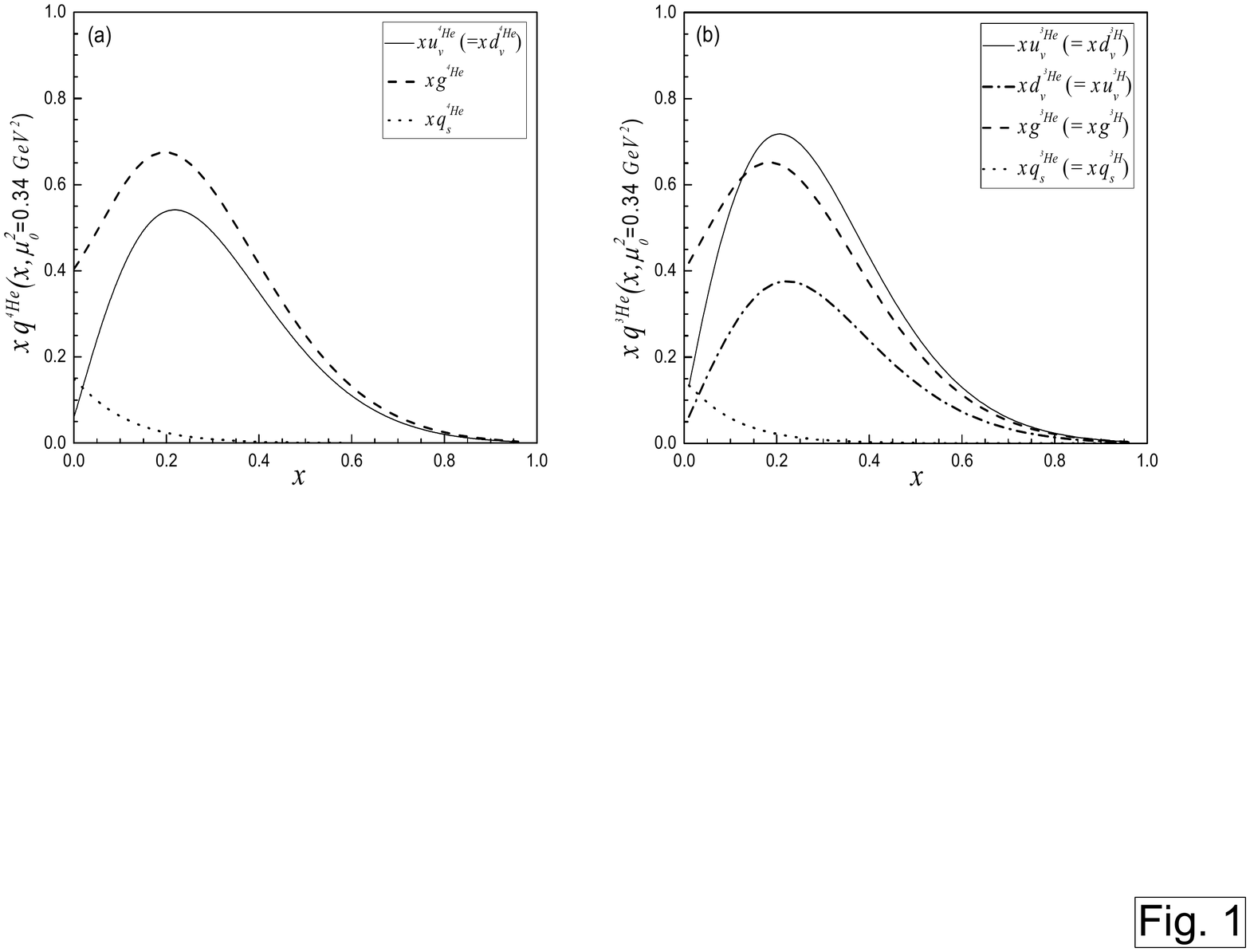}
\caption{ Panel (a): PDFs of $^4He$ nucleus versus $x$, for $(m_a,
\epsilon_0^a)$ pair of (320, 120 $MeV$)  ($a$ = ${\mathcal{U}}$,
${\mathcal{D}}$) and $b$ = 0.8 $fm$ at the hadronic scale, $\mu_0^2$
= 0.34 $GeV^2$. The dash curve represents the gluon distribution,
while the solid and dotted curves indicate the valence and sea quark
distributions, respectively. Panel (b):  PDFs of $^3He$ nucleus
versus $x$, for ($m_{\mathcal{U}}$, $\epsilon_0^{\mathcal{U}}$) and
($m_{\mathcal{D}}$, $\epsilon_0^{\mathcal{D}}$) pairs of  (300, 130
MeV) and (325, 115 MeV), respectively, and $b$ = 0.8 $fm$ at the
hadronic scale, $\mu_0^2$ = 0.34 $GeV^2$.  The solid and dotted-dash
curves are the valence up (down) and down (up) quark distributions
of $^3He$ ($^3H$) nucleus, respectively, while the dotted and dash
curves indicate the  sea quark and gluon distributions,
respectively.}
\end{figure}
\begin{figure}[h!]
  \includegraphics [width=\linewidth]{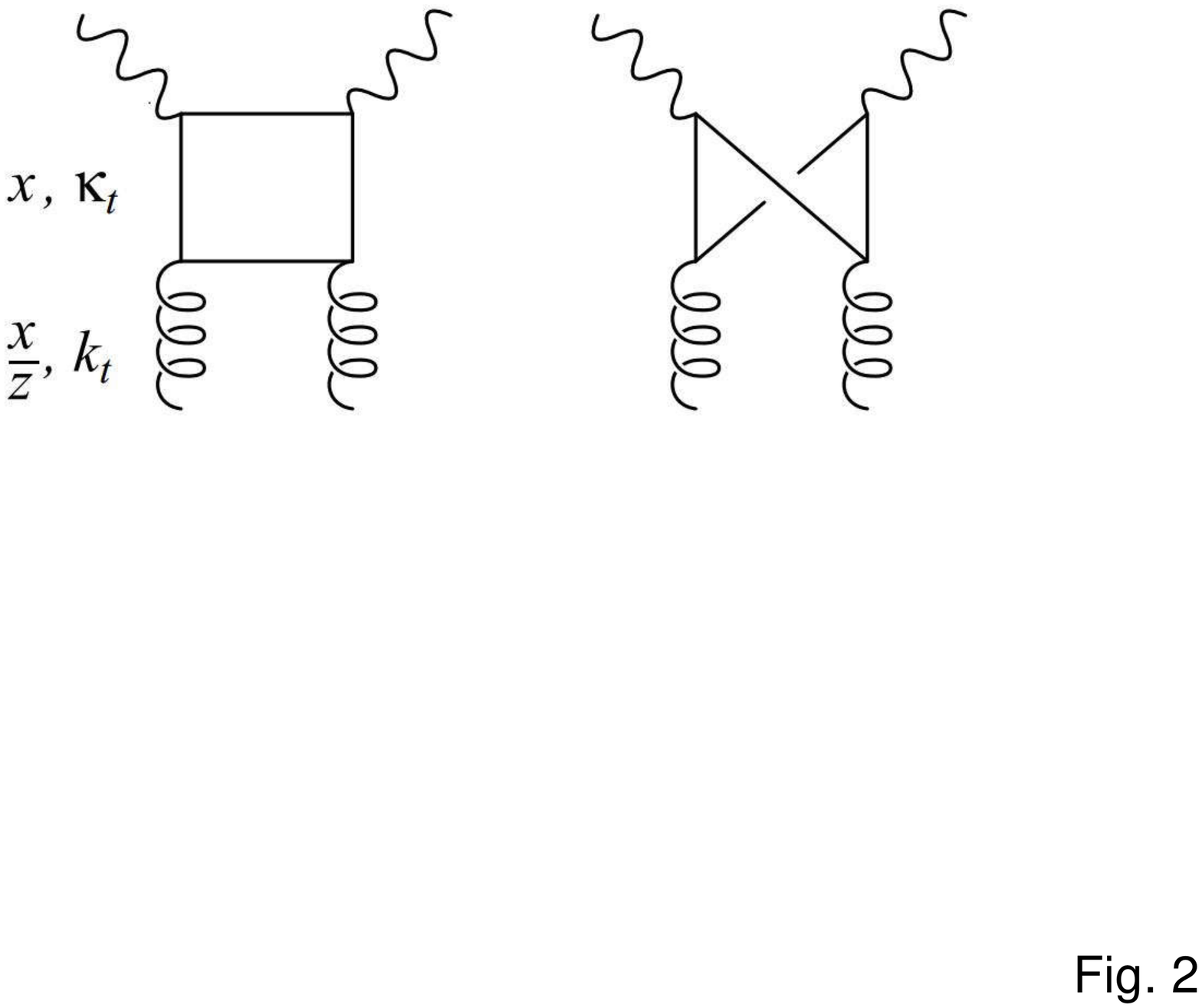}
\caption{The quark box and crossed-box diagrams which demonstrate
the contribution of the unintegrated gluon distributions,
$f_g$($x$/$z$, $k_t^2$, $\mu^2$), to the structure function, $F_2$.}
\end{figure}
\begin{figure}[h!]
  \includegraphics [width=\linewidth]{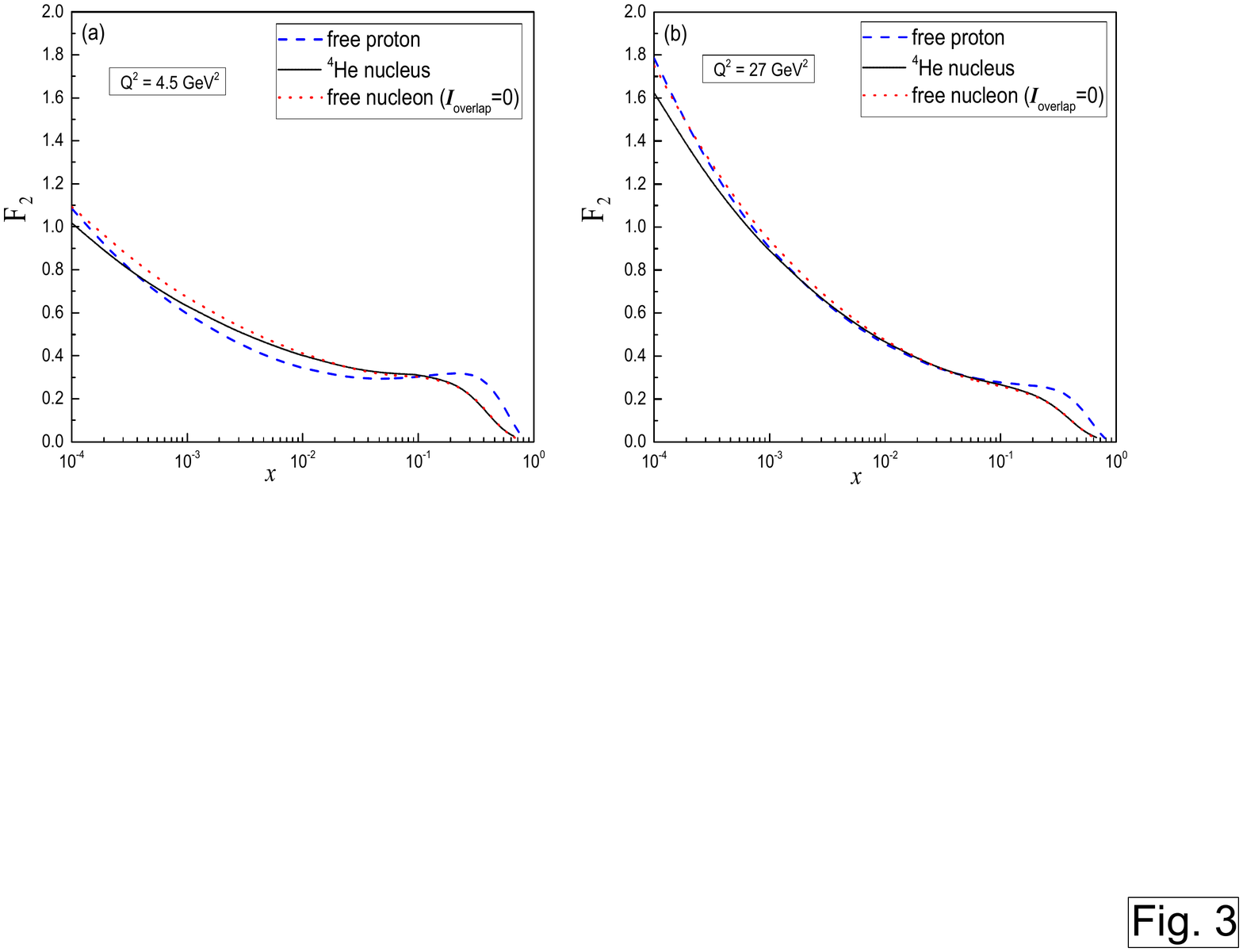}
\caption{The comparison of the SFs of the $^4He$ nucleus in the KMR
approach (the full curves) with those of the free proton using the
MSTW-2008 data sets as inputs (the dash curves), at the energy
scales 4.5 $GeV^2$ (panel (a)),  and 27 $GeV^2$ (panel (b)). the
dotted curves indicate our hypothetical free nucleon (by setting the
overlap integral $\mathcal{I}$ equal to zero in the momentum density
formula  of $^4He$ nucleus) SFs in the KMR approach. All SFs are
calculated with considering the three lightest quark flavors ($u$,
$d$, $s$).}
\end{figure}
\begin{figure}[h!]
  \includegraphics [width=\linewidth]{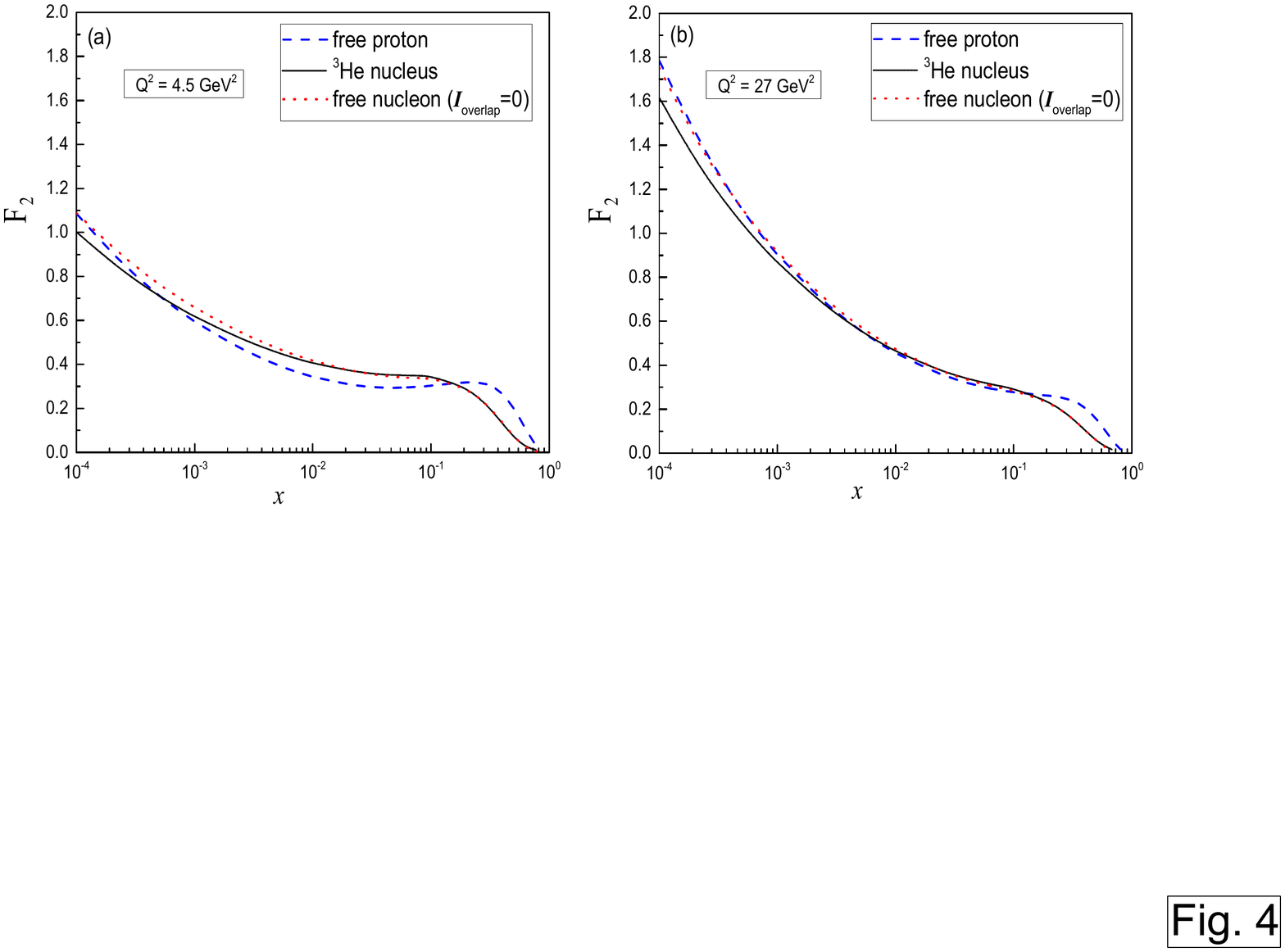}
\caption{The same as the figure 3, but for the $^3He$ nucleus.}
\end{figure}
\begin{figure}[h!]
  \includegraphics [width=\linewidth]{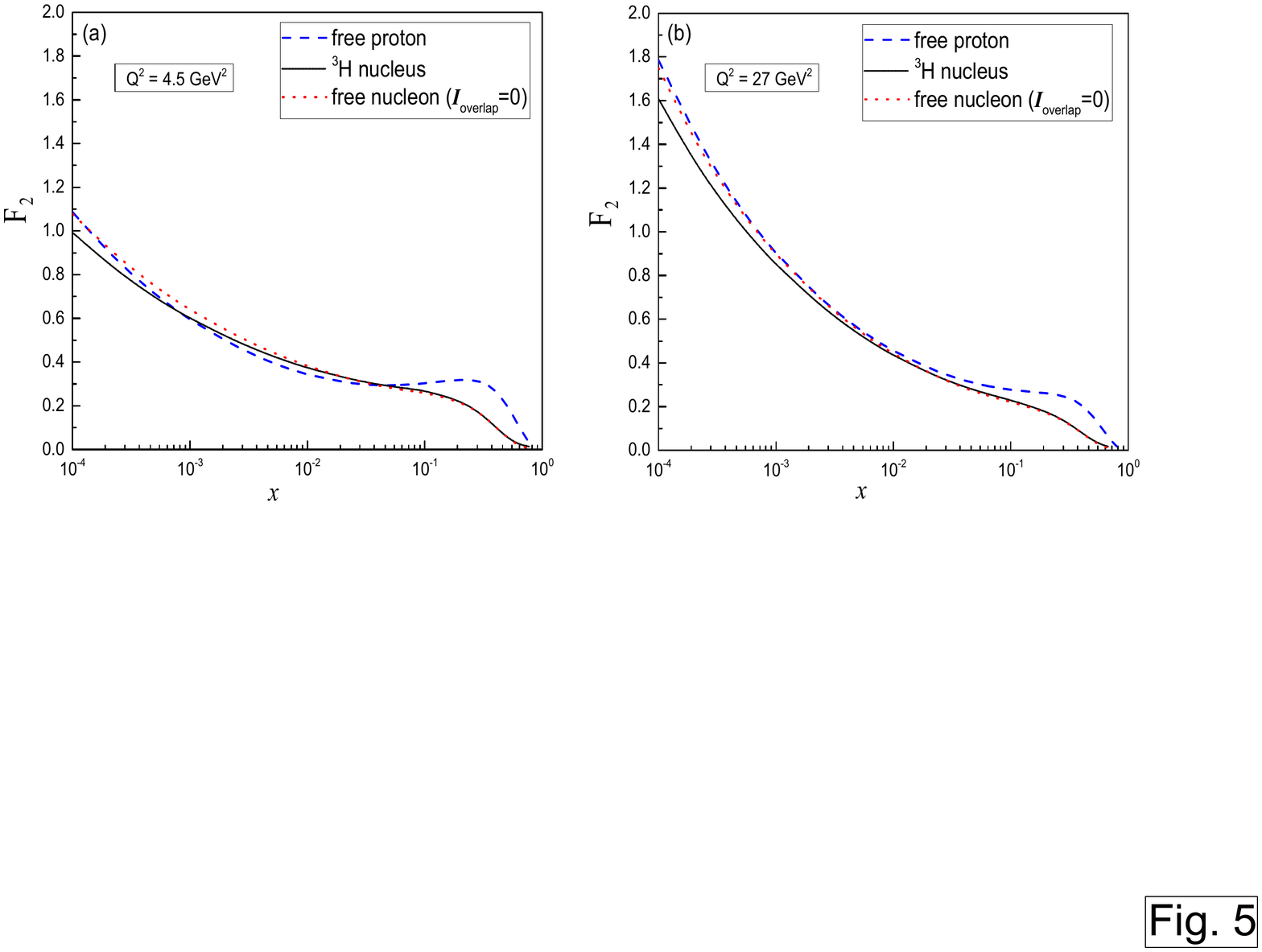}
\caption{The same as the figure 3, but for the $^3H$ nucleus.}
\end{figure}
\begin{figure}[h!]
  \includegraphics [ scale=0.6]{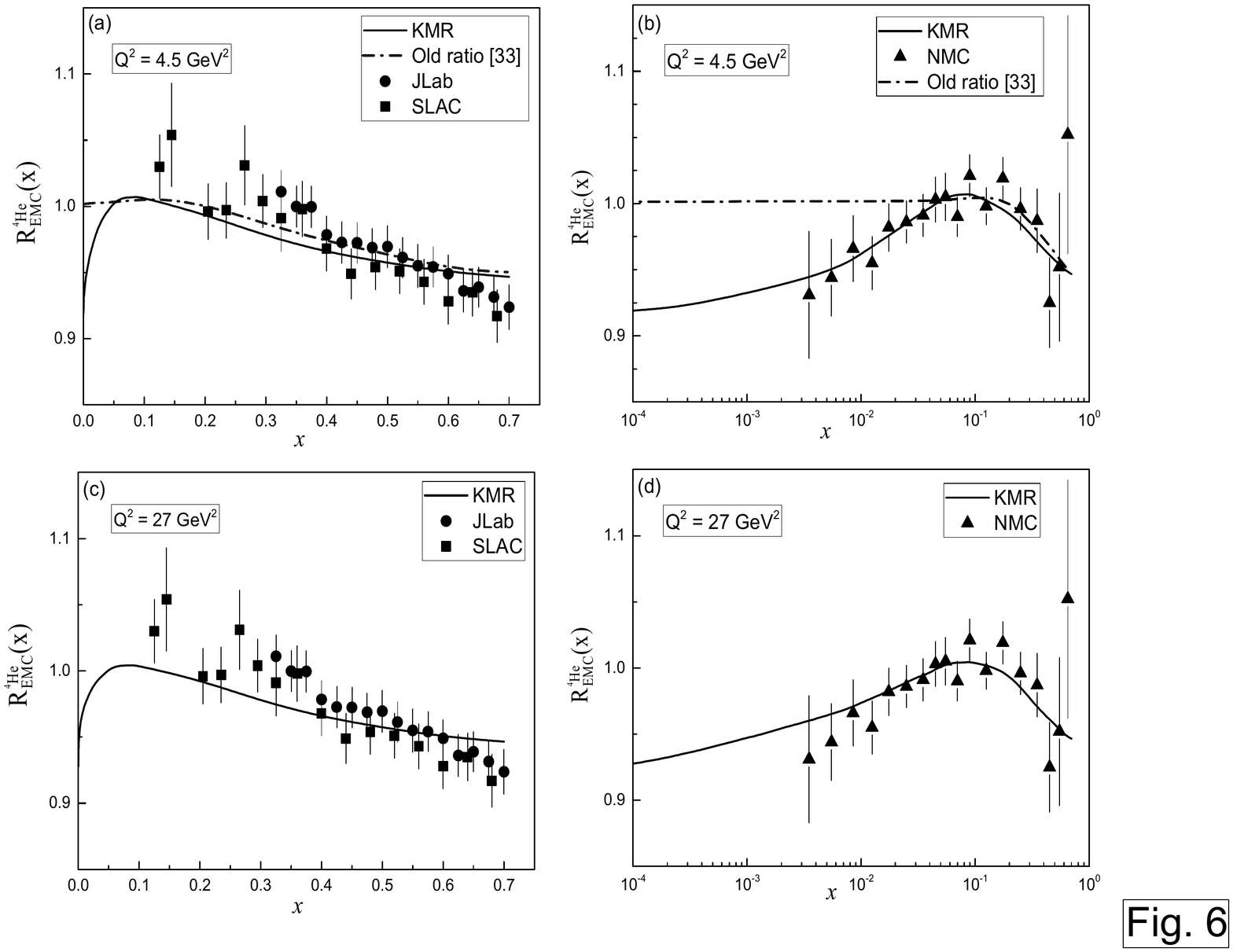}
\caption{The EMC ratio of $^4He$ nucleus in the $k_t$-factoization
framework by using the KMR UPDFs as inputs (the full curves),  at
the energy scales 4.5 $GeV^2$ (panels (a) and (b)),  and 27 $GeV^2$
(panels (c) and (d)). The circles, the triangles, and the squares
are from JLab \cite{Seely,Malace}, NMC \cite{New,Malace}, and SLAC
\cite{Gomez,Malace} experimental data, respectively.  the
dotted-dash curves in the panels (a) and (b), are given from
reference \cite{Hadian3} at $b$ = 0.8 $fm$ and $Q^2$ = 0.34 $GeV^2$,
in which the contributions of UPDFs are not accounted in the $^4He$
EMC calculations. }
\end{figure}
\begin{figure}[h!]
  \includegraphics [width=\linewidth]{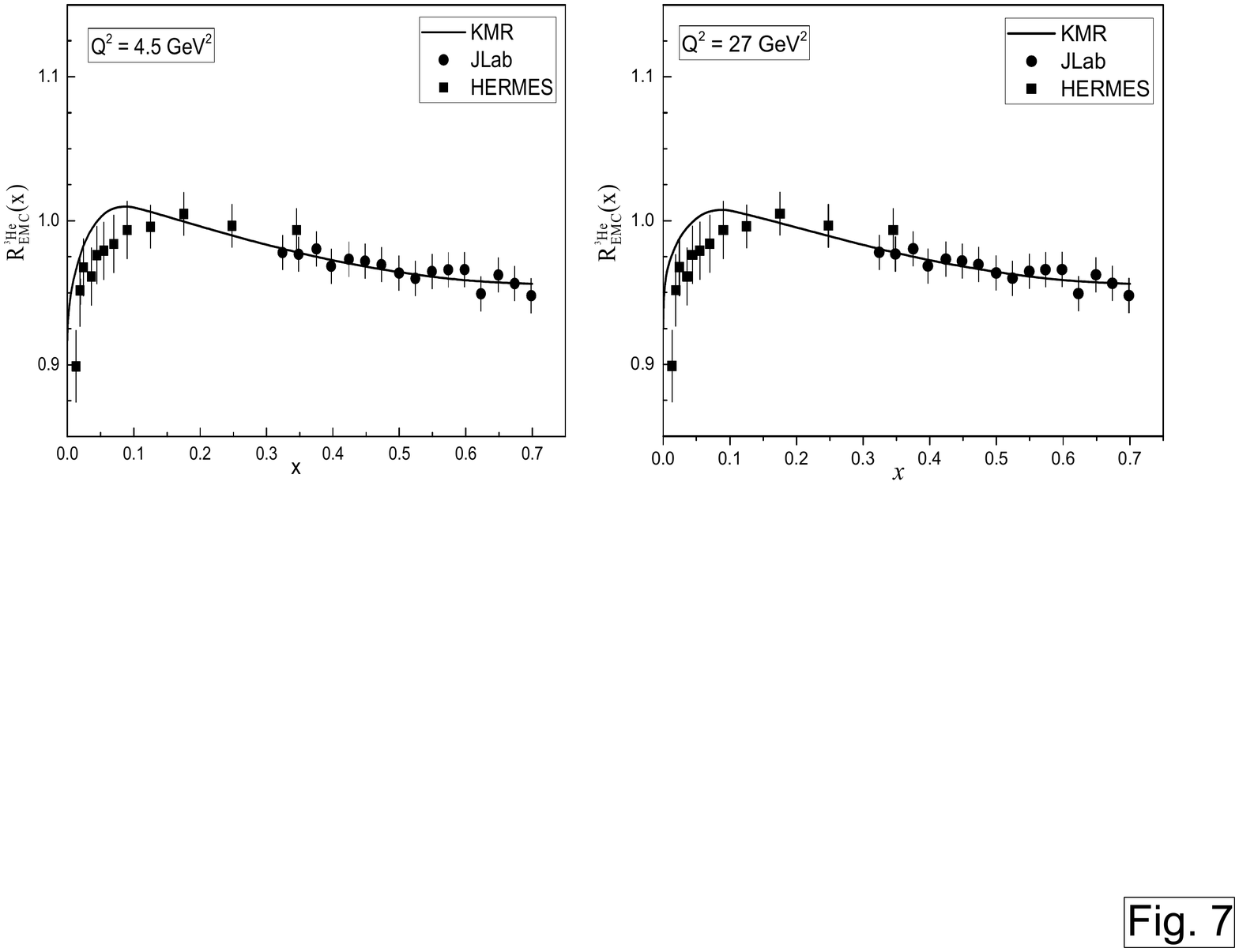}
\caption{The EMC ratio of $^3He$ nucleus in the $k_t$-factoization
framework by using the KMR UPDFs as inputs (the full curves),  at
the energy scales 4.5 $GeV^2$ (left panel),  and 27 $GeV^2$ (right
panel). The filled circles and the filled squares are the
experimental data from JLab \cite{Seely,Malace} and HERMES
\cite{Malace,Airapetisn}, respectively.}
\end{figure}
\begin{figure}[h!]
  \includegraphics [ width=\linewidth]{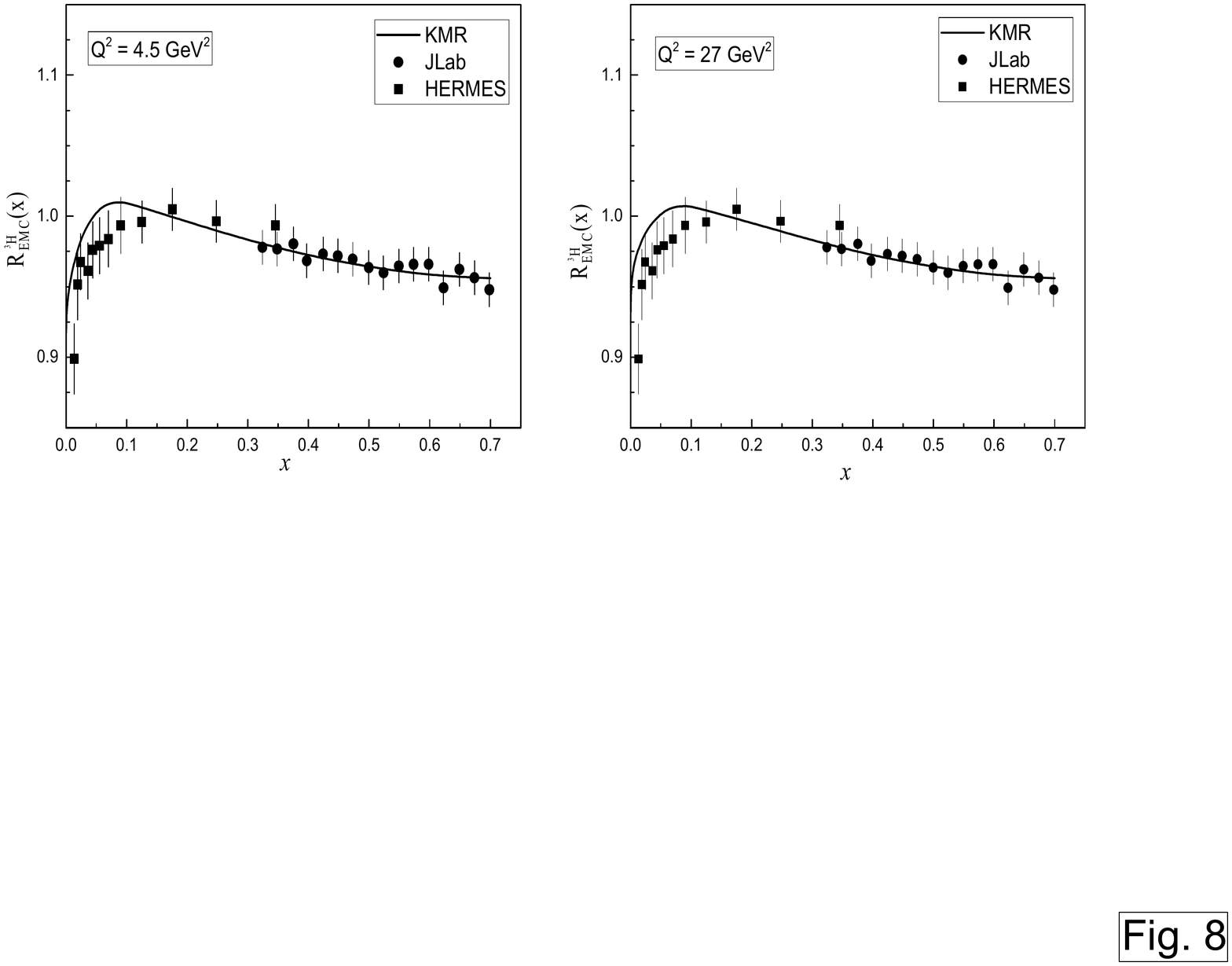}
\caption{The same as the figure 7, but for the $^3H$ nucleus.}
\end{figure}
\begin{figure}[h!]
  \includegraphics [ width=\linewidth]{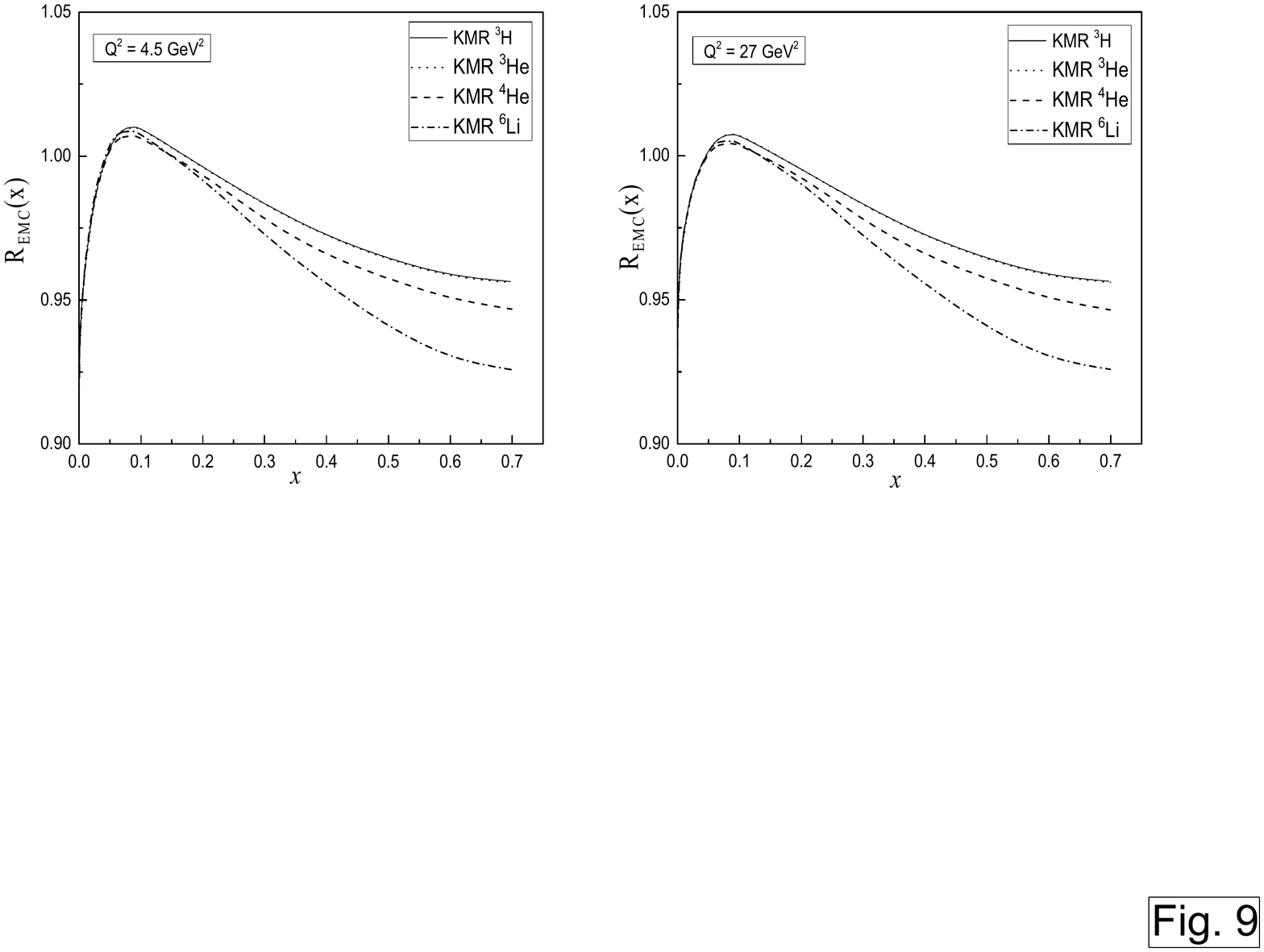}
\caption{The comparisons of EMC ratios of $^6Li$ (the dash-dotted
curves), $^4He$ (the dash curves), $^3He$ (the dotted curves) and
$^3H$ (the full curves) nuclei in the KMR approach at the energy
scales 4.5 $GeV^2$ (left panel) and 27 $GeV^2$ (right panel).  The
$^6Li$  EMC ratios are given from the reference \cite{Hadian4}.}
\end{figure}
\end{document}